\documentclass[american,aps,prl,twocolumn]{revtex4-1}
\usepackage[T1]{fontenc}
\usepackage[latin9]{inputenc}
\usepackage{amsmath}
\usepackage{graphicx}
\usepackage{babel}
\begin{document}

\title{Generalized curvature modified plasma dispersion functions and Dupree
renormalization of toroidal ITG}

\author{Ö. Gültekin$^{1}$, Ö. D. Gürcan$^{2,3}$}
\email{ozgur.gurcan@lpp.polytechnique.fr}

\selectlanguage{american}%

\affiliation{$^{1}$Department of Physics, Istanbul University, Istanbul, 34134,
Turkey}

\affiliation{$^{2}$CNRS, Laboratoire de Physique des Plasmas, Ecole Polytechnique,
Palaiseau}

\affiliation{$^{3}$ Sorbonne Universités, UPMC Univ Paris 06, Paris}
\begin{abstract}
A new generalization of curvature modified plasma dispersion functions
is introduced in order to express Dupree renormalized dispersion relations
used in quasi-linear theory. For instance the Dupree renormalized
dispersion relation for gyrokinetic, toroidal ion temperature gradient
driven (ITG) modes, where the Dupree's diffusion coefficient is assumed
to be a low order polynomial of the velocity, can be written entirely
using generalized curvature modified plasma dispersion functions:
$K_{nm}$'s. Using those, Dupree's formulation of renormalized quasi-linear
theory is revisited for the toroidal ITG mode. The Dupree diffusion
coefficient has been obtained as a function of velocity using an iteration
scheme, first by assuming that the diffusion coefficient is constant
at each $v$ (i.e. applicable for slow dependence), and then substituting
the resulting $v$ dependence in the form of complex polynomial coefficients
into the $K_{nm}$'s for verification. The algorithm generally converges
rapidly after only a few iterations. Since the quasi-linear calculation
relies on an assumed form for the wave-number spectrum, especially
around its peak, practical usefullness of the method is to be determined in actual
applications.
\end{abstract}
\maketitle

\section{Introduction}

\subsection{Background}

Most linear dispersion relations, relevant for kinetic waves in plasmas
are described using combinations and/or generalizations of plasma
dispersion functions\cite{fried:1961}. Many generalizations and modifications
were proposed in the past\cite{robinson:89,summers:91,hellberg:02,xie:13} and efficiencies of the methods are compared\cite{xie:17}.

A particular such generalization is the introduction of the so called
curvature modified plasma dispersion functions. These functions, which
appear naturally in the formulation of the local, linear dispersion
relation of the toroidal ion temperature gradient driven (ITG) mode,
can be used to write the linear dispersion relations of various drift
instabilities including ITG or its homologue the electron temperature
gradient driven (ETG) mode in a generic form. Since the analytical
continuation of these functions have already been incorporated in
their definitions, they can be used in order to study damped modes
as well as the usual unstable branches.

In this paper we introduce a further generalization of the curvature
modified plasma dispersion functions which allows us to include a
Dupree diffusion coefficient that has a velocity dependence in the
form of a second order polynomial. This allows us to describe the
perturbed or renormalized dispersion relations for toroidal drift
instabilities using these functions. 

The algorithm that we use in this paper is as follows. We start complex
frequencies that are obtained as the solutions of the linear dispersion
relation $\omega_{k}=\omega_{rk}+i\gamma_{k}$. Since the Dupree
diffusion coefficient $D_{D}$ is usually defined in terms of itself
via a complicated equation of the sort $D_{D}\sim\sum_{k}D_{0k}/\left(i\omega_{rk}+\gamma_{k}+D_{D}\right)$,
we first compute it by taking it as zero on the right hand side. This
gives us the first iteration $D_{D}^{\left(1\right)}$. Using $D_{D}=D_{D}^{\left(1\right)}$
in the perturbed dispersion equation$\varepsilon\left(\omega,\boldsymbol{k},D_{D}^{\left(1\right)}\left(v\right)\right)=0$,
we compute the dominant renormalized complex frequency, which can
be dubbed $\omega_{k}^{\left(1\right)}$. Using $D_{D}^{\left(1\right)}$
and $\omega_{k}^{\left(1\right)}$ we compute $D_{D}^{\left(2\right)}$
and so on, until the difference between $\left(D_{D}^{\left(n\right)}-D_{D}^{\left(n-1\right)}\right)/D_{D}^{\left(n-1\right)}$
is smaller than a predefined tolerance. The algorithm converges rapidly
in only 3-4 iterations in most cases.

The paper is organized as follows.

\subsection{Curvature modified dispersion functions}

The functions, dubbed $I_{nm}\left(\zeta_{\alpha},\zeta_{\beta},b\right)$,
and defined for $Im\left[\zeta_{\alpha}\right]>0$ as: 
\begin{align}
 & I_{nm}\left(\zeta_{\alpha},\zeta_{\beta},b\right)\equiv\nonumber \\
 & \frac{2}{\sqrt{\pi}}\int_{0}^{\infty}dx_{\perp}\int_{-\infty}^{\infty}dx_{\parallel}\frac{x_{\perp}^{n}x_{\parallel}^{m}J_{0}^{2}\left(\sqrt{2b}x_{\perp}\right)e^{-x^{2}}}{\left(x_{\parallel}^{2}+\frac{x_{\perp}^{2}}{2}+\zeta_{\alpha}-\zeta_{\beta}x_{\parallel}\right)}\;\text{,}\label{eq:db_int}
\end{align}
can be written as a 1D integral of a combination of plasma dispersion
functions as discussed in detail in Ref. \cite{gurcan:14}:

\begin{align}
 & I_{nm}\left(\zeta_{\alpha},\zeta_{\beta},b\right)=\nonumber \\
 & \int_{0}^{\infty}s^{\frac{n-1}{2}}G_{m}\left(z_{1}\left(s\right),z_{2}\left(s\right)\right)J_{0}\left(\sqrt{2bs}\right)^{2}e^{-s}ds\;\mbox{,}\nonumber \\
 & \quad(Im\left[\zeta_{\alpha}\right]>0)\label{eq:inm1}
\end{align}
using the straightforward multi-variable generalization of the standard
plasma dispersion function: $G_{m}\left(z_{1},z_{2},\cdots,z_{n}\right)\equiv\frac{1}{\sqrt{\pi}}\int_{-\infty}^{\infty}\frac{x^{m}e^{-x^{2}}}{\prod_{i=1}^{n}\left(x-z_{i}\right)}dx$
with
\begin{equation}
z_{1,2}\left(s\right)=\frac{1}{2}\left(\zeta_{\beta}\pm\sqrt{\zeta_{\beta}^{2}-2\left(s+2\zeta_{\alpha}\right)}\right)\;\mbox{.}\label{eq:z12}
\end{equation}
The $G_{m}\left(z_{1},z_{2}\right)$ in (\ref{eq:inm1}) can be written
in terms of the standard plasma dispersion function as:

\begin{align}
G_{m}\left(z_{1},z_{2}\right) & =\frac{1}{\sqrt{\pi}\left(z_{1}-z_{2}\right)}\bigg[z_{1}^{m}Z_{0}\left(z_{1}\right)-z_{2}^{m}Z_{0}\left(z_{2}\right)\nonumber \\
 & +\sum_{k=2}^{m}\left(z_{1}^{k-1}-z_{2}^{k-1}\right)\Gamma\left(\frac{m-k+1}{2}\right)\bigg]\;,\label{eq:gm}
\end{align}
which can be implemented using the Weideman method \cite{weideman:94}.

Furthermore, in order to extend the validity of the definition (\ref{eq:inm1})
we must use $I_{nm}=I_{nm}^{'}+\Delta I_{nm}$ {[}where $I_{nm}^{'}$
is the integral in (\ref{eq:inm1}){]}, where \begin{widetext}
\begin{align}
\Delta I_{nm}\left(\zeta_{\alpha},\zeta_{\beta},b\right)=-i\sqrt{\pi}2^{\frac{\left(n+3\right)}{2}}w^{\frac{n}{2}}\int_{-1}^{1}d\mu & \left(1-\mu^{2}\right)^{\frac{\left(n-1\right)}{2}}\left(\mu\sqrt{w}+\frac{\zeta_{\beta}}{2}\right)^{m}\nonumber \\
 & J_{0}^{2}\left(2\sqrt{b\left(1-\mu^{2}\right)w}\right)e^{-2\left(1-\mu^{2}\right)w-\left(\mu\sqrt{w}+\frac{\zeta_{\beta}}{2}\right)^{2}}\times\begin{cases}
0 & \zeta_{\alpha i}>0\quad\mbox{or }w_{r}<0\\
\frac{1}{2} & \zeta_{\alpha i}=0\quad\mbox{and }w_{r}>0\\
1 & \zeta_{\alpha i}<0\quad\mbox{and }w_{r}>0
\end{cases}\label{eq:resid}
\end{align}
\end{widetext} with $w=\frac{\zeta_{\beta}^{2}}{4}-\zeta_{\alpha}$,
$\zeta_{\alpha i}=\text{Im}\left(\zeta_{\alpha}\right)$ and $w_{r}=\text{Re}\left[w\right]$. 

Derivatives of curvature modified dispersion functions can also be
defined as analytical functions and implemented similarly \cite{gultekin:18}.

\subsection{Toroidal ITG}

Following the kinetic formulation of toroidal ITG mode\cite{kim:94,kuroda:98}
starting from the linear gyrokinetic equation\cite{catto:78,frieman:82,hahm:88},
which can be written in Fourier space for the non-adiabatic part of
the distribution function as:
\begin{align}
\bigg(\frac{\partial}{\partial t}+ & iv_{\parallel}k_{\parallel}+i\hat{\omega}_{D}\left(v\right)\bigg)\delta g_{\boldsymbol{k}}\nonumber \\
= & \left(\frac{\partial}{\partial t}+i\omega_{*T}\left(v\right)\right)\frac{e}{T_{i}}\delta\Phi_{\boldsymbol{k}}F_{0}J_{0}\left(\frac{k_{\perp}v_{\perp}}{\Omega_{i}}\right)\label{eq:gk}
\end{align}
Where $\hat{\omega}_{D}\left(v\right)=\frac{\omega_{D}}{2}\left(\frac{v_{\parallel}^{2}}{v_{ti}^{2}}+\frac{v_{\perp}^{2}}{2v_{ti}^{2}}\right)$,
$\omega_{*Ti}\left(v\right)\equiv\omega_{*i}\left[1+\left(\frac{v^{2}}{2v_{ti}^{2}}-\frac{3}{2}\right)\eta_{i}\right]$,
$\omega_{D}=2\frac{L_{n}}{R}\omega_{*i}$, $\omega_{*i}=v_{ti}\rho_{i}k_{y}/L_{n}$,
$L_{n}=-\left(d\ln n_{0}/dr\right)^{-1}$, $\eta_{i}=d\ln T_{i}/d\ln n_{0}$,
$R$ is the major radius of the tokamak, $v_{ti}$ is the ion thermal
velocity, $\Omega_{i}$ is the ion cyclotron frequency and $\rho_{i}$
is the ion Larmor radius. The gyrokinetic equation is complemented
by the quasi-neutrality relation, which is written here for adiabatic
electrons for $k_{\parallel}\neq0$:
\begin{equation}
\frac{e}{T_{e}}\delta\Phi_{k}=-\frac{e}{T_{i}}\delta\Phi_{k}+\int J_{0}\delta g_{k}d^{3}v\;\text{.}\label{eq:qn}
\end{equation}
Taking the Laplace-Fourier transform of (\ref{eq:gk}), solving for
$\delta g_{\mathbf{k},\omega}$ and substituting the result into (\ref{eq:qn}),
we obtain the following dispersion relation, written in terms of $\varepsilon\left(\omega,\mathbf{k}\right)$,
the plasma dielectric function:

\begin{equation}
\varepsilon\left(\omega,\mathbf{k}\right)\equiv1+\frac{1}{\tau}-P\left(\omega,\mathbf{k}\right)=0\;\text{.}\label{eq:drel-1}
\end{equation}
where
\begin{align*}
 & P\left(\omega,\mathbf{k}\right)=\\
 & \frac{1}{\sqrt{2\pi}v_{ti}^{3}}\int\frac{\left(\omega-\omega_{*Ti}\left(v\right)\right)J_{0}^{2}\left(\frac{k_{\perp}v_{\perp}}{\Omega_{i}}\right)e^{-\frac{v^{2}}{2v_{ti}^{2}}}}{\left(\omega-v_{\parallel}k_{\parallel}-\omega_{Di}\frac{1}{2}\left(\frac{v_{\parallel}^{2}}{v_{ti}^{2}}+\frac{v_{\perp}^{2}}{2v_{ti}^{2}}\right)\right)}v_{\perp}dv_{\perp}dv_{\parallel}
\end{align*}
ormalising $k$ with $\rho_{i}^{-1}$ and Using $\omega/\left|k_{y}\right|\rightarrow\omega$,
and $\omega_{D}/\left|k_{y}\right|\rightarrow\omega_{D}$, the dispersion
relation (\ref{eq:drel-1}) can be written as:

\begin{align}
\varepsilon\left(\omega,\mathbf{k}\right)\equiv & 1+\frac{1}{\tau}+\frac{1}{\omega_{Di}}\bigg(I_{10}\left[\omega+\left(1-\frac{3}{2}\eta_{i}\right)\right]\nonumber \\
 & +\left(I_{30}+I_{12}\right)\eta_{i}\bigg)=0\label{eq:eps_pdf}
\end{align}
where $I_{nm}\equiv I_{nm}\left(-\frac{\omega}{\omega_{Di}},-\frac{\sqrt{2}k_{\parallel}}{\omega_{Di}k_{y}},b\right)$
with $b\equiv k_{\perp}^{2}$. 

\section{Generalized curvature modified dispersion functions}

A further generalization of $I_{nm}$'s can be proposed in the form
of the following integrals:
\begin{align}
 & K_{nm}\left(\zeta_{a},\zeta_{b},\zeta_{c},\zeta_{d},b\right)=\nonumber \\
 & \frac{2}{\sqrt{\pi}}\int_{0}^{\infty}dx_{\perp}\int dx_{\parallel}\frac{x_{\perp}^{n}x_{\parallel}^{m}J_{0}^{2}\left(\sqrt{2b}x_{\perp}\right)e^{-x^{2}}}{\left(x_{\parallel}^{2}+\zeta_{a}+\zeta_{b}x_{\parallel}+\zeta_{c}x_{\perp}+\zeta_{d}x_{\perp}^{2}\right)}\label{eq:knm}
\end{align}
with complex variables $\zeta_{a}$, $\zeta_{b}$, $\zeta_{c}$ and
$\zeta_{d}$, and the real variable $b$, so that, we can write
\[
I_{nm}\left(\zeta_{\alpha},\zeta_{\beta},b\right)=K_{nm}\left(\zeta_{\alpha},-\zeta_{\beta},0,\frac{1}{2},b\right)\;\text{.}
\]
The definition of the $I_{nm}$'s in terms of of the generalized plasma
dispersion functions $G_{m}\left(z_{1},z_{2},\cdots,z_{n}\right)$
as in (\ref{eq:inm1}) can be extended to $K_{nm}$'s:
\begin{align*}
 & K_{nm}\left(\zeta_{a},\zeta_{b},\zeta_{c},\zeta_{d},b\right)=\\
 & 2\int_{0}^{\infty}x_{\perp}^{n}G_{m}\left(z_{1}^{'}\left(x_{\perp}\right),z_{2}^{'}\left(x_{\perp}\right)\right)J_{0}\left(\sqrt{2b}x_{\perp}\right)^{2}e^{-x_{\perp}^{2}}dx_{\perp}\;\mbox{,}
\end{align*}
using 
\[
z_{1,2}^{'}\left(x_{\perp}\right)=\frac{1}{2}\left(-\zeta_{b}\mp\sqrt{\zeta_{b}^{2}-4\left(\zeta_{c}x_{\perp}+\zeta_{d}x_{\perp}^{2}+\zeta_{a}\right)}\right)\;\mbox{.}
\]
Note that the function defined in (\ref{eq:knm}) can represent any
complex polynomial denominator up to second order in $x_{\parallel}$
and $x_{\perp}$. The resulting functions can be implemented using
(\ref{eq:gm}) to write the $G_{m}$'s and computing the remaining
one dimensional $x_{\perp}$ integral using a numerical quadrature.

\subsection{Analytical Continuation}

The integral in (\ref{eq:knm}) has a singularity in the complex plane
when the polynomial in the denominator vanishes. Since the polynomial
is a quadratic one that is a function of both $x_{\parallel}$ and
$x_{\perp}$, it vanishes on a curve. If the coefficients were all
real, this would be a curve in two dimensions. However since the coefficients
are complex, the actual curve lives in 4 dimensional space made of
$x_{\parallel r}$, $x_{\parallel i}$, $x_{\perp r}$ and $x_{\perp i}$
(i.e. real and imaginary parts of the extended $x_{\parallel}$ and
$x_{\perp}$). However since it remains a ``curve'', we can still
parametrize it.

The analytical continuation of the $I_{nm}$'s are performed by first
scaling the $x_{\perp}$ and $x_{\parallel}$ so that the ellipse
becomes a circle, and then switching to polar coordinates so that
the singularity becomes a point on the $r$ axis. The residue contribution
is then equal to the integral over the angular variable on the circle.
This residue is to be added to the original integral when $\text{Im}\left(\zeta_{\alpha}\right)<0$
and $w_{r}>0$ as shown in (\ref{eq:resid}).

In the case of the $K_{nm}$ the issue is more complicated. Considering
the denominator
\[
d\left(x_{\perp},x_{\parallel}\right)=x_{\parallel}^{2}+\zeta_{b}x_{\parallel}+\zeta_{c}x_{\perp}+\zeta_{d}x_{\perp}^{2}+\zeta_{\alpha}
\]
can be written using $\rho=\sqrt{\left(x_{\parallel}+\frac{\zeta_{br}}{2}\right)^{2}+\left(\sqrt{\zeta_{dr}}x_{\perp}\right)^{2}}$
and $\mu=\left(x_{\parallel}+\frac{\zeta_{br}}{2}\right)/\rho$ as:
\begin{align*}
d\left(\rho,\mu\right)= & \rho^{2}\left(1+i\frac{\zeta_{di}\left(1-\mu^{2}\right)}{\zeta_{dr}}\right)\\
 & +\left(\zeta_{c}\frac{\sqrt{\left(1-\mu^{2}\right)}}{\sqrt{\zeta_{dr}}}+i\zeta_{bi}\mu\right)\rho-w
\end{align*}
where $w=\frac{\zeta_{br}^{2}}{4}+i\frac{\zeta_{bi}\zeta_{br}}{2}-\zeta_{\alpha}$.
The denominator becomes zero at the two complex roots:
\begin{equation}
\rho_{\pm}=-\frac{\beta}{2\alpha}\pm\frac{1}{2a}\sqrt{4\alpha w+\beta^{2}}\label{eq:phopm}
\end{equation}
with $\alpha\equiv\left(1+i\frac{\zeta_{di}\left(1-\mu^{2}\right)}{\zeta_{dr}}\right)$
and $\beta\equiv\left(\zeta_{c}\frac{\sqrt{\left(1-\mu^{2}\right)}}{\sqrt{\zeta_{dr}}}+i\zeta_{bi}\mu\right)$.

Note that when we do the transformation from $x_{\perp}$, $x_{\parallel}$
to $\rho$ and $\mu$, the surface element becomes:
\[
dx_{\perp}dx_{\parallel}=\frac{\rho}{\sqrt{\zeta_{dr}\left(1-\mu^{2}\right)}}d\mu d\rho
\]
so that the integral can be written as: 
\begin{align*}
 & K_{nm}\left(\zeta_{a},\zeta_{b},\zeta_{c},\zeta_{d},b\right)\equiv\\
 & \frac{2}{\sqrt{\pi}}\int_{-1}^{1}d\mu\int_{0}^{\infty}\rho d\rho\frac{x_{\perp}\left(\rho,\mu\right)^{n}x_{\parallel}\left(\rho,\mu\right)^{m}\left(J_{0}^{\rho}\right)^{2}e^{-x^{2}\left(\rho,\mu\right)}}{\alpha\left(\rho-\rho_{+}\right)\left(\rho-\rho_{-}\right)\sqrt{\zeta_{dr}\left(1-\mu^{2}\right)}}
\end{align*}
where the shorthand notation $J_{0}^{\rho}=J_{0}\left(\sqrt{2b}x_{\perp}\left(\rho,\mu\right)\right)$
and $x^{2}\left(\rho,\mu\right)=x_{\perp}\left(\rho,\mu\right)^{2}+x_{\parallel}\left(\rho,\mu\right)^{2}$
has been used. This means that the residue contribution
\begin{align*}
 & \Delta K_{nm}\left(\zeta_{a},\zeta_{b},\zeta_{c},\zeta_{d},b\right)\equiv\\
 & 4\sqrt{\pi}i\int_{-1}^{1}\rho_{+}d\mu\frac{x_{\perp}\left(\rho_{+},\mu\right)^{n}x_{\parallel}\left(\rho_{+},\mu\right)^{m}\left(J_{0}^{\rho_{+}}\right)^{2}e^{-x\left(\rho_{+},\mu\right)^{2}}}{\sqrt{4\alpha w+\beta^{2}}\sqrt{\zeta_{dr}\left(1-\mu^{2}\right)}}
\end{align*}
 should be added to the integral when $\text{Im}\left[\rho_{+}\left(\mu\right)\right]>0$
(and we should add $1/2$ times the residue contribution when $\text{Im}\left[\rho_{+}\left(\mu\right)\right]=0$).
Note that writing the condition $\text{Im}\left[\rho_{+}\left(\mu\right)\right]>0$
in terms of real and imaginary parts of $\zeta$ parameters is complicated
enough that we find it more practical to check this condition by computing
$\rho_{+}$ using (\ref{eq:phopm}) numerically.

\begin{figure}
\includegraphics[width=0.5\columnwidth]{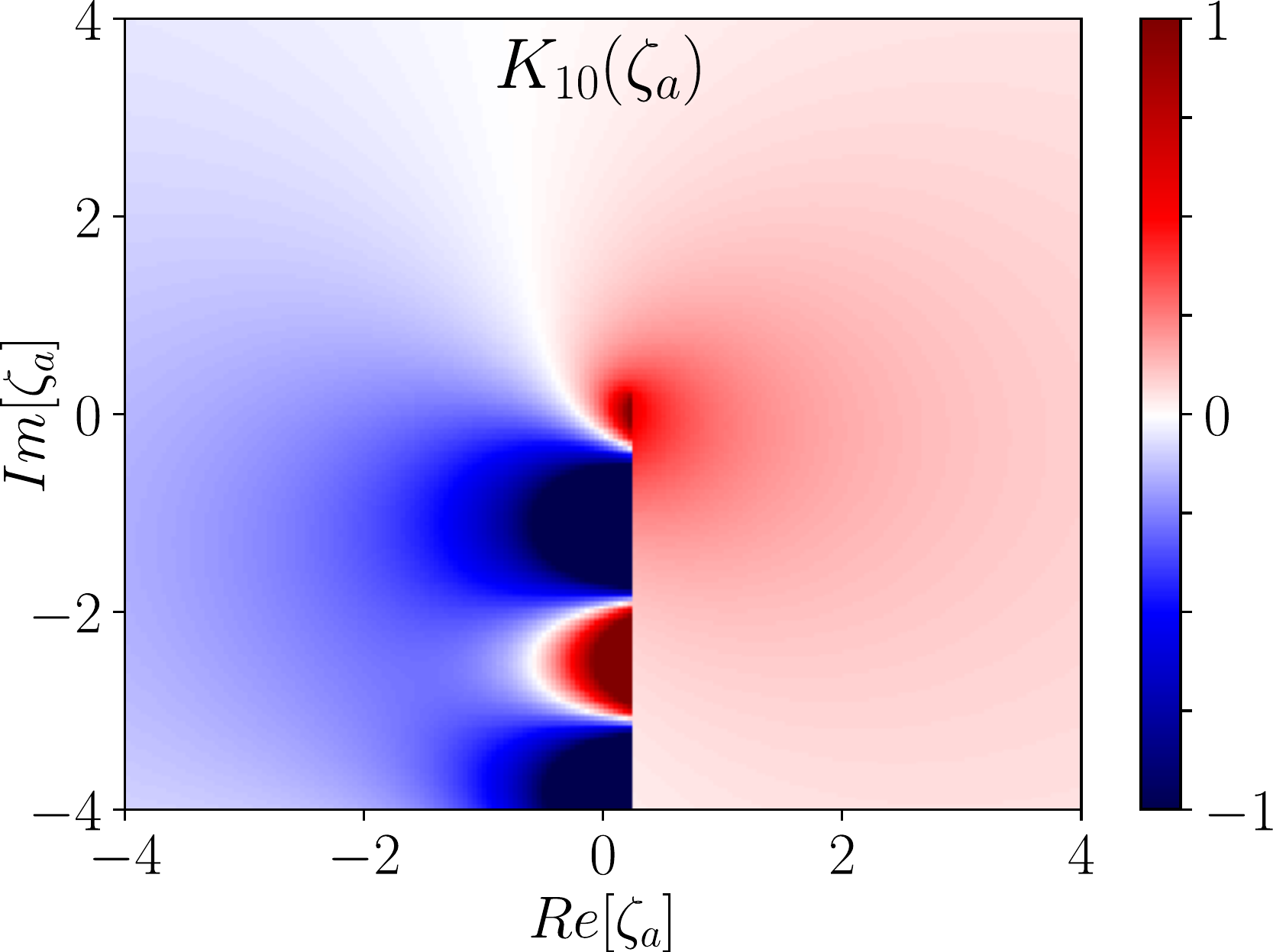}\includegraphics[width=0.5\columnwidth]{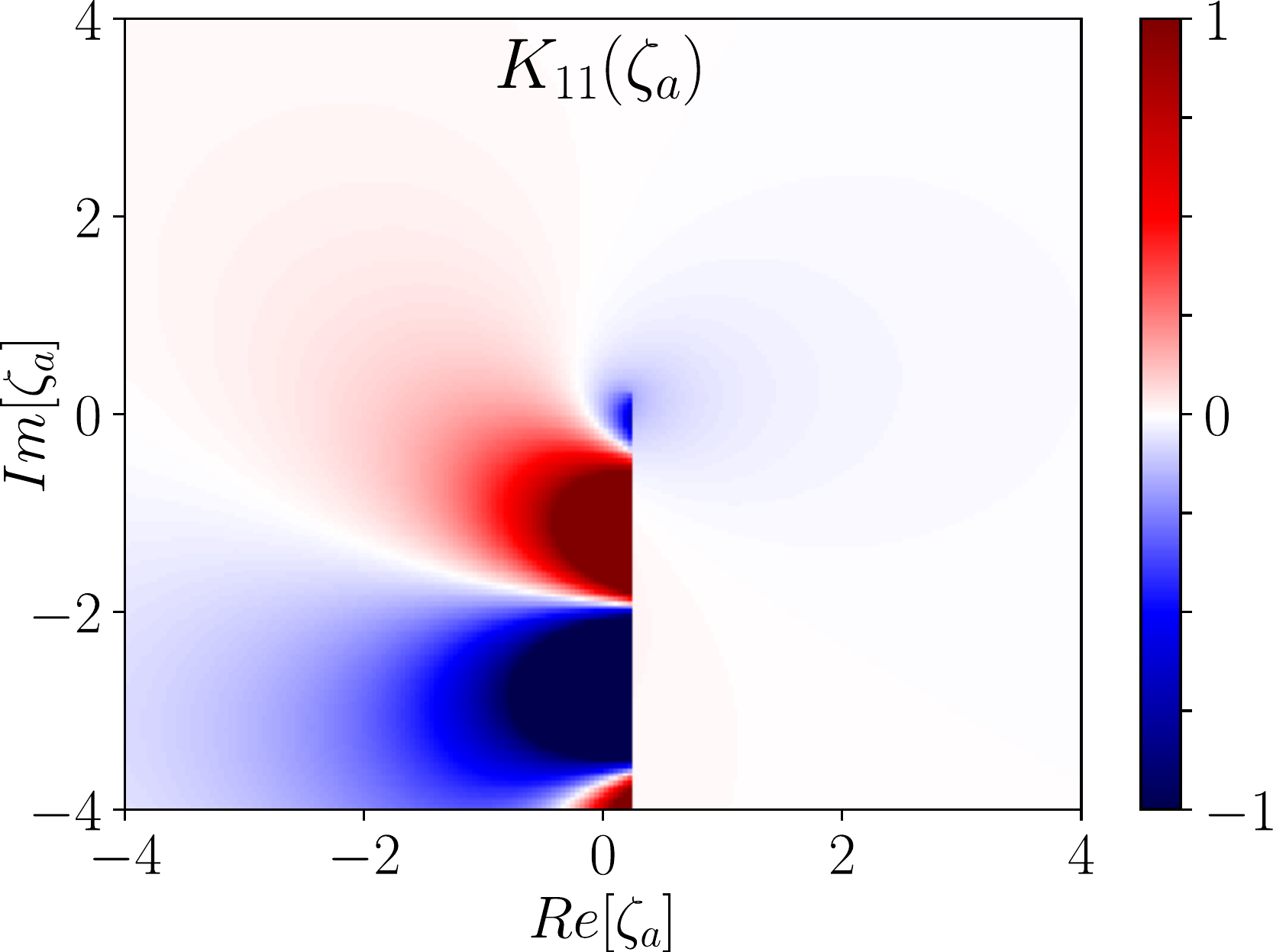}

\includegraphics[width=0.5\columnwidth]{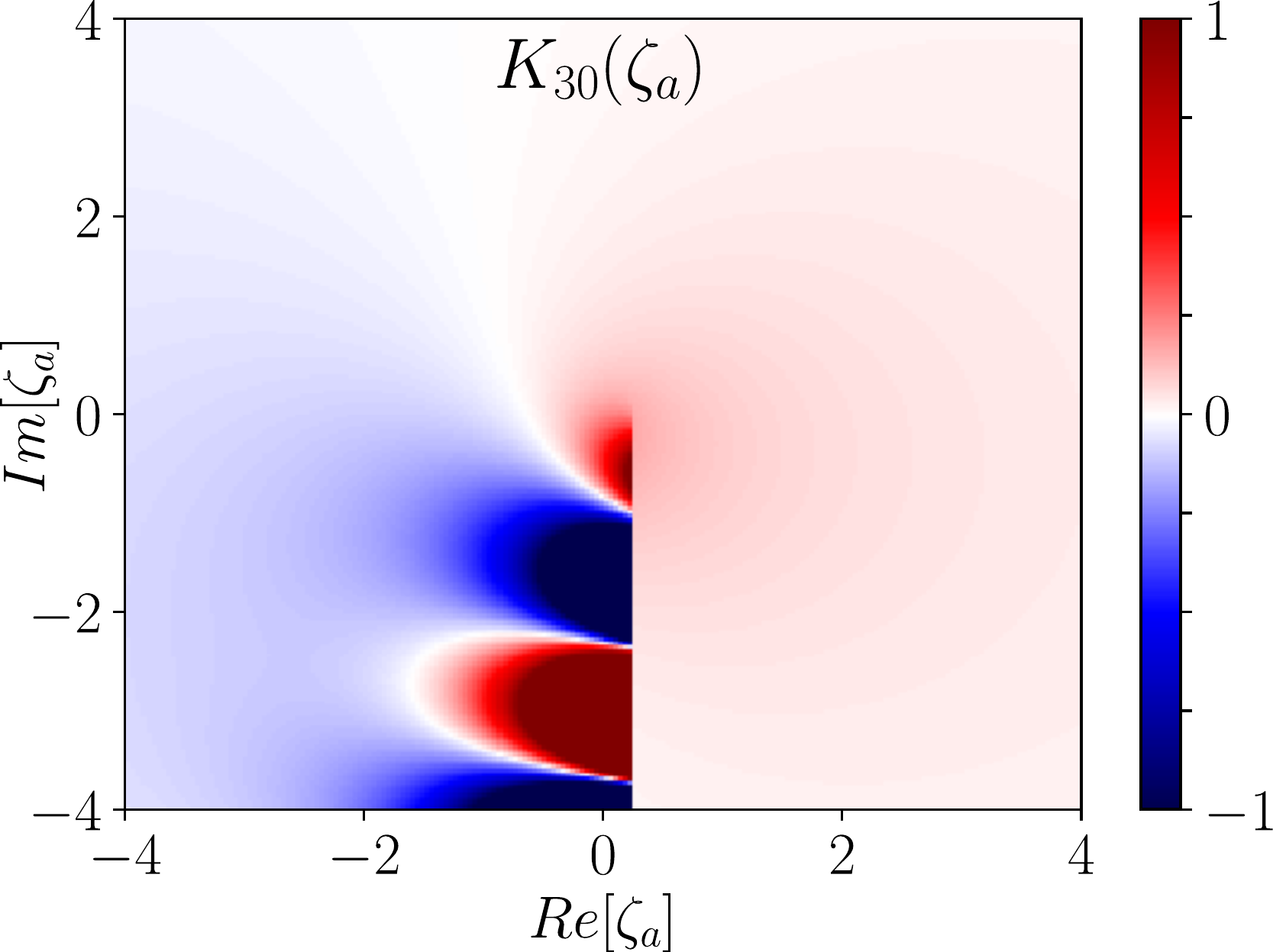}\includegraphics[width=0.5\columnwidth]{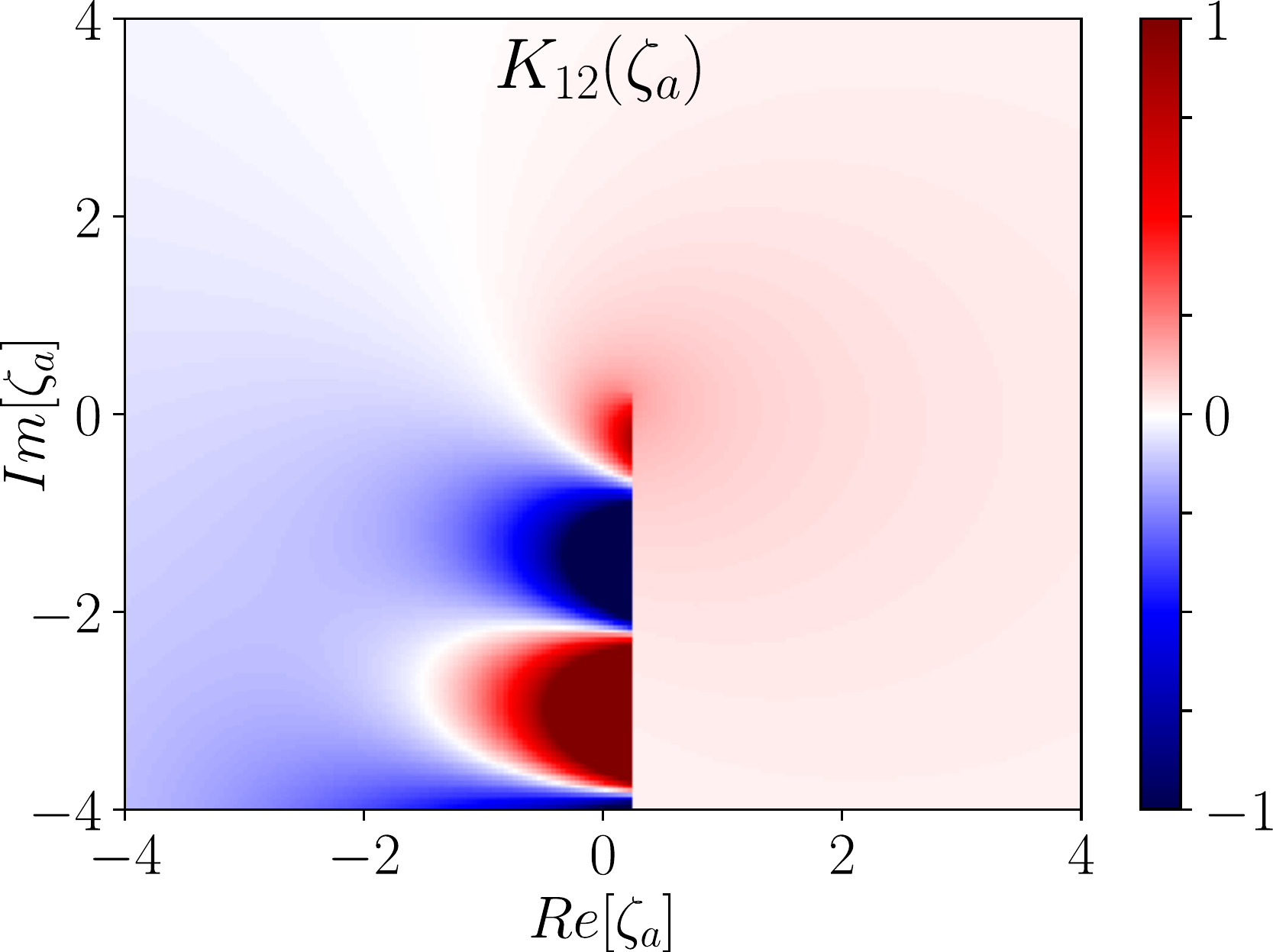}

\caption{\label{fig:knms_set1}Plots of $K_{10}$, $K_{11}$, $K_{30}$ and
$K_{12}$ (from top left to bottom right) as functions of the real
and imaginary parts of $\zeta_{a}$, where the other parameters are
fixed at $\zeta_{b}=1.0+0.5i$, $\zeta_{c}=0.5+0.5i$, $\zeta_{d}=0.5+0.2i$
and $b=1.0$.}
\end{figure}

\begin{figure}
\includegraphics[width=0.5\columnwidth]{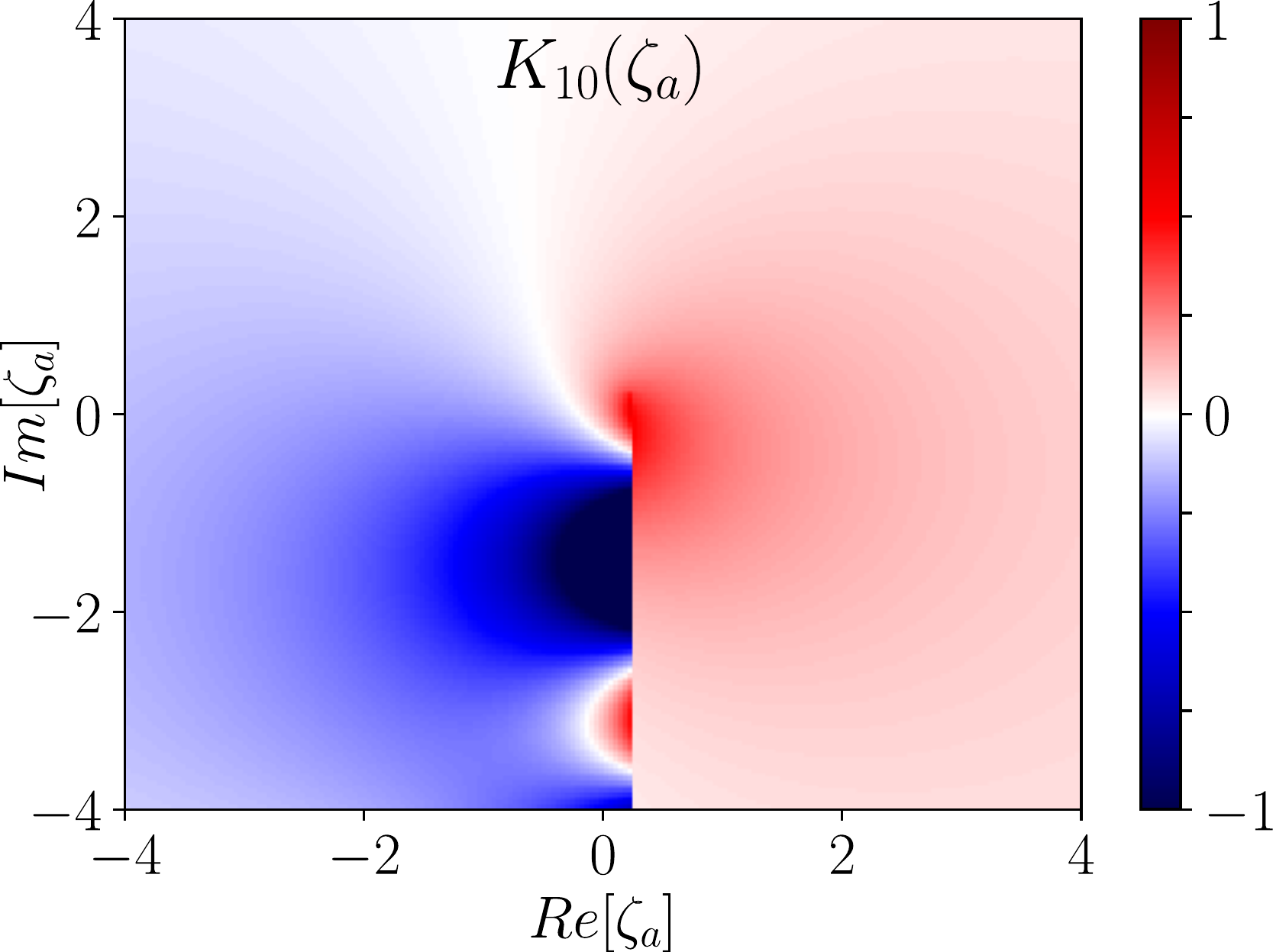}\includegraphics[width=0.5\columnwidth]{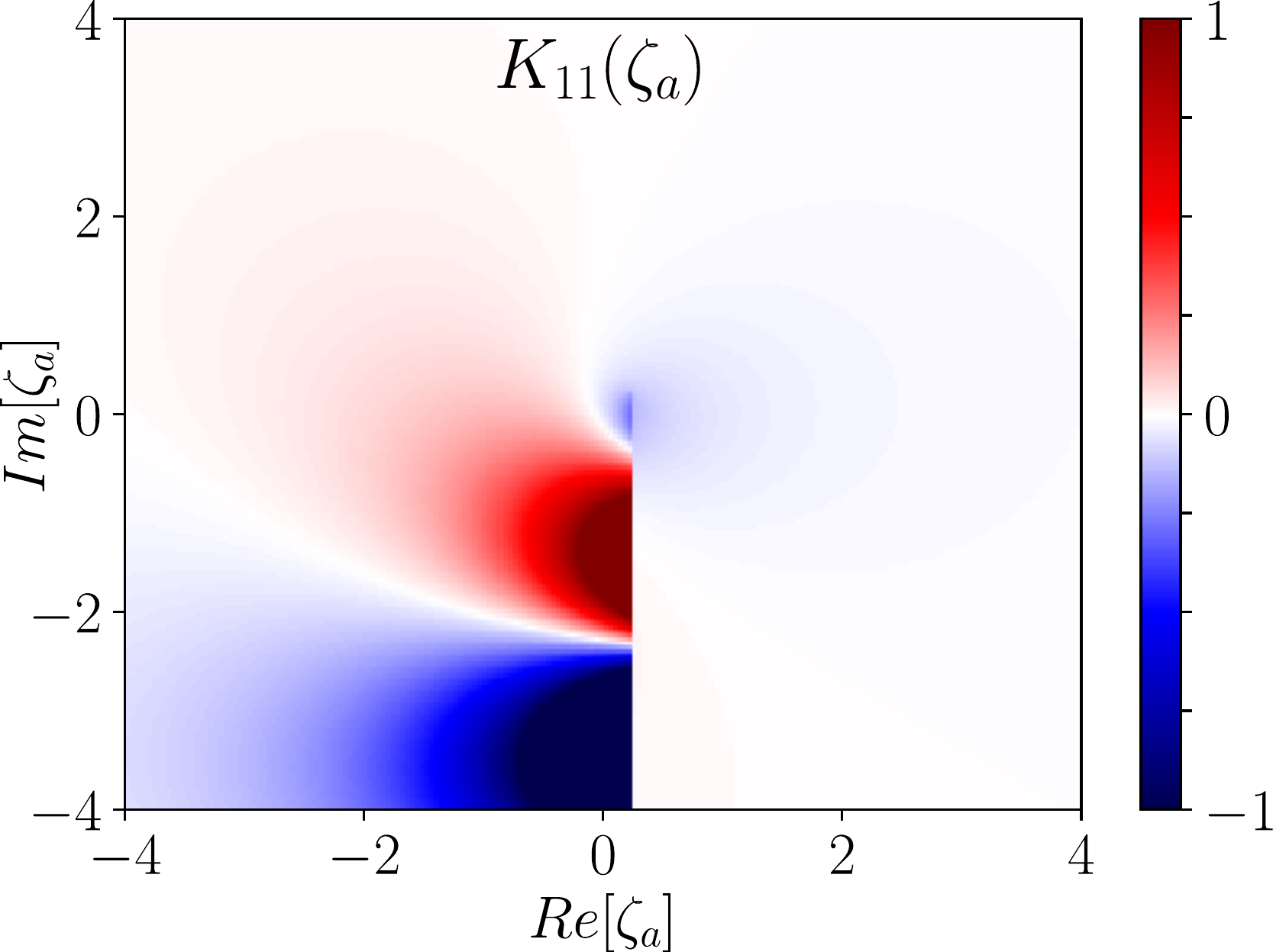}

\includegraphics[width=0.5\columnwidth]{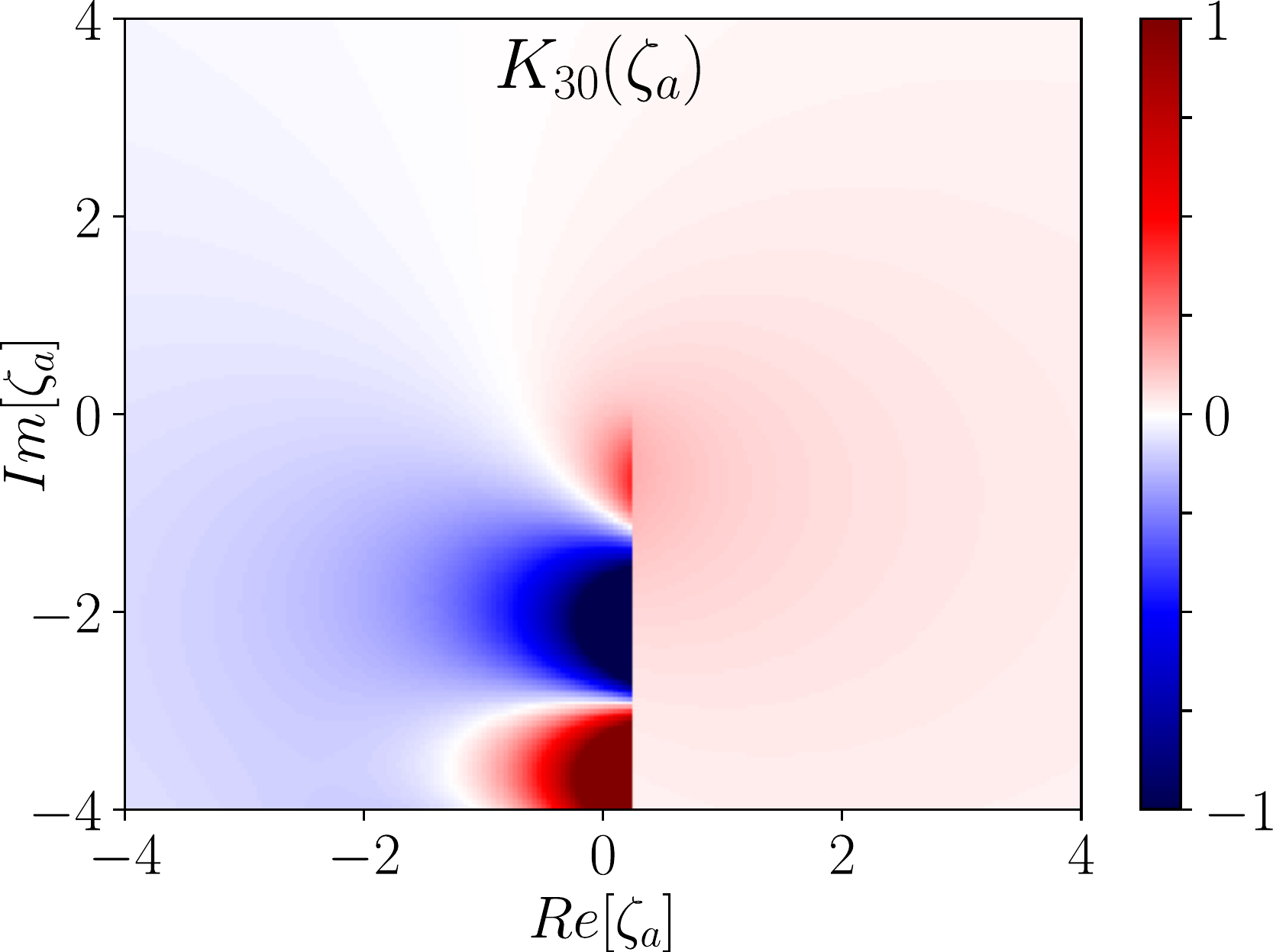}\includegraphics[width=0.5\columnwidth]{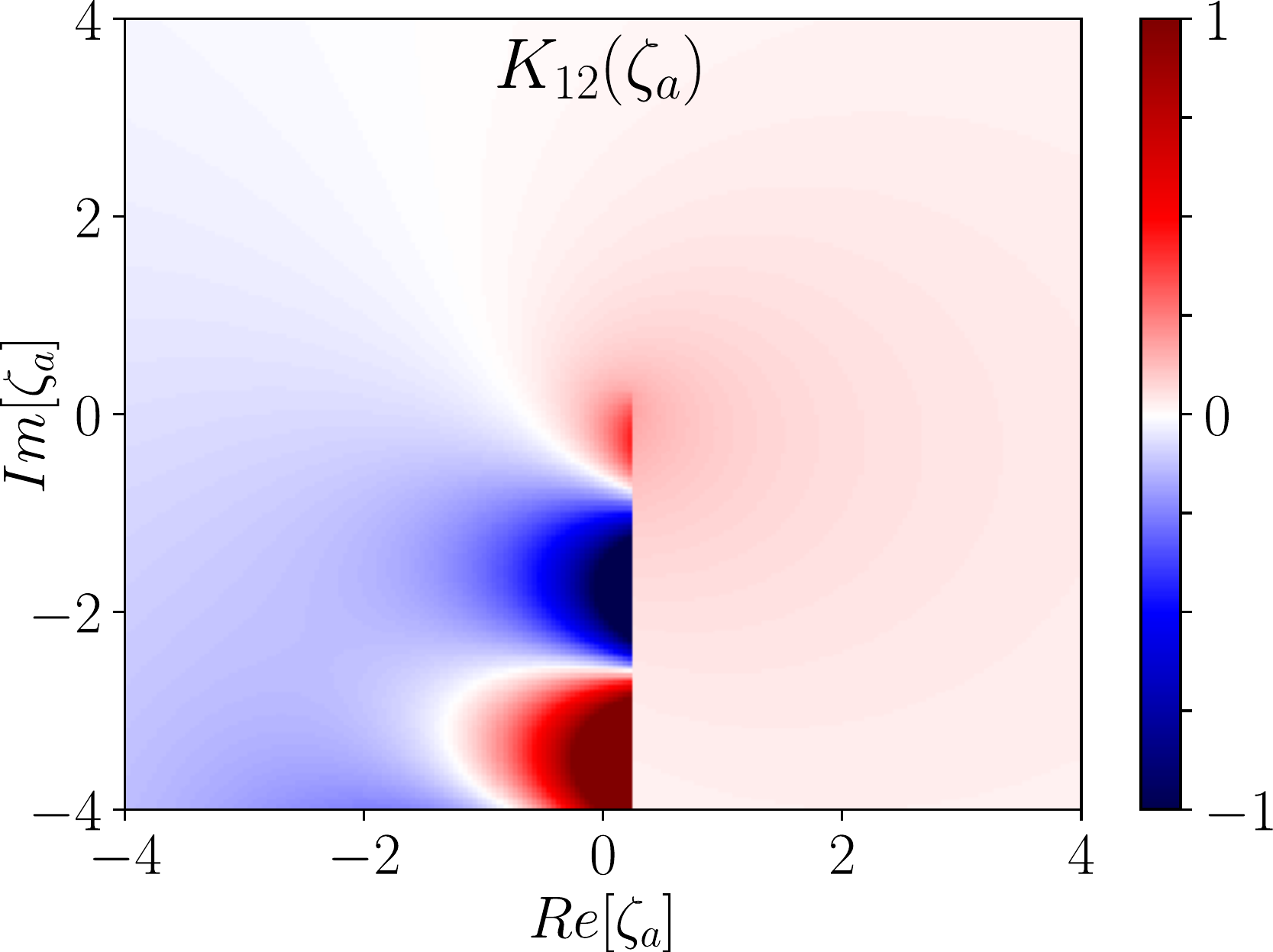}

\caption{\label{fig:knms_set3}Plots of $K_{10}$, $K_{11}$, $K_{30}$ and
$K_{12}$ (from top left to bottom right) as functions of the real
and imaginary parts of $\zeta_{a}$, where the other parameters are
fixed at the same parameters as figure \ref{fig:knms_set1} but with
$\zeta_{c}=0.5+1.0i$.}
\end{figure}

As a function of the real and imaginary parts of its primary variable
$\zeta_{a}$, the function $K_{nm}$ can be plotted fixing the values
of its other variables. The results are shown in figures \ref{fig:knms_set1}
and \ref{fig:knms_set3}.

\section{Dupree Renormalization}

Dupree's renormalized form of the gyrokinetic equation can be written
as\cite{balescu:book:anom,krommes:02}:
\begin{align}
\bigg(\frac{\partial}{\partial t}+ & iv_{\parallel}k_{\parallel}+i\hat{\omega}_{D}\left(v\right)+D_{D}\left(v\right)k_{\perp}^{2}\bigg)\delta g_{\boldsymbol{k}}\nonumber \\
= & \left(\frac{\partial}{\partial t}+i\omega_{*T}\left(v\right)+D_{D}\left(v\right)k_{\perp}^{2}\right)\frac{e}{T_{i}}\delta\Phi_{\boldsymbol{k}}F_{0}J_{0}\left(\frac{k_{\perp}v_{\perp}}{\Omega_{i}}\right)\;\text{.}\label{eq:gkren}
\end{align}
Assuming a general complex $D_{D}\left(v\right)$, and short correlation
times, we can write the discrete equivalent of Dupree's integral equation
as:

\begin{equation}
D_{D}\left(v_{\perp},v_{\parallel}\right)=\sum_{k}\frac{k^{2}J_{0k}^{2}\left|\delta\Phi_{k}\right|^{2}}{i\left(\omega_{k}^{*}-\omega_{Dk}\left(v\right)-k_{\parallel}v_{\parallel}\right)+D_{D}^{*}\left(v_{\perp},v_{\parallel}\right)k_{\perp}^{2}}\label{eq:dupre_int}
\end{equation}
Note that here $\omega_{k}=\omega_{rk}+i\gamma_{k}$ is the solution
of the \emph{renormalized} linear dispersion relation due to (\ref{eq:gkren}):
\begin{align}
 & \varepsilon\left(\omega,\mathbf{k}\right)\equiv1+\frac{1}{\tau}\nonumber \\
 & -\frac{1}{\sqrt{2\pi}v_{ti}^{3}}\int\frac{\left(\omega-\omega_{*Ti}\left(v\right)+iD_{D}\left(v\right)k_{\perp}^{2}\right)J_{0k}^{2}e^{-\frac{v^{2}}{2v_{ti}^{2}}}}{\omega-v_{\parallel}k_{\parallel}-\hat{\omega}_{Di}\left(v\right)+iD_{D}\left(v\right)k_{\perp}^{2}}v_{\perp}dv_{\perp}dv_{\parallel}\label{eq:drel2}
\end{align}
which is usually considered to introduce a nonlinear damping on top
of the linear solution (i.e. $\omega_{k}=\omega_{k,\text{lin}}-i\eta_{\text{eddy}}$).
This is exactly the case, if the Dupree diffusion coefficient is a
real constant independent of $v$. However in the case of a complex
function of $v_{\perp}$ and $v_{\parallel}$, the issue is more complicated.

One approach to this problem is to consider the dependence of $D_{D}$
on $v$ as being weak. This allows us to solve (\ref{eq:dupre_int})
at each $v$ seperately while considering $D_{D}$ as a constant in
solving $\omega_{k}$ from (\ref{eq:drel2}). Of course this is only
an intermediate step and one should finally consider a $v$ dependent
$D_{D}\left(v\right)$ and somehow substitute that into (\ref{eq:drel2})
and show that $\omega_{k}$ that one can obtain together with the
$D_{D}\left(v\right)$ gives us back the $D_{D}\left(v\right)$ that
we used as a requirement of verification of this solution.

Unfortunately for an arbitrary function of $v$, this is hard to do.
However when $D_{D}\left(v\right)$ is computed at each $v$ by assuming
it as a constant in (\ref{eq:drel2}), the function that is obtained
as a function of $v_{\parallel}$ or $v_{\perp}$ is rather close
to a low order polynomial in a range of $v$ values. It means that
we can fit a polynomial of the form
\begin{equation}
D_{D}=d_{0}+d_{1}x_{\parallel}+d_{2}x_{\parallel}^{2}+d_{3}x_{\perp}+d_{4}x_{\perp}^{2}\label{eq:poly}
\end{equation}
where $x_{\perp,\parallel}=v_{\perp,\parallel}/\sqrt{2}v_{ti}$, to
the form obtained by solving (\ref{eq:dupre_int}) at each $v_{\perp}$
and $v_{\parallel}$. For simplicity we performed this fit by fixing
$v_{\perp}=0$ and computing $D\left(v_{\parallel},0\right)$ and
fitting and then fixing $v_{\parallel}=0$ and computing $D_{D}\left(0,v_{\perp}\right)$
and fitting.

For a given $v_{\perp}$, $v_{\parallel}$ we solve (\ref{eq:dupre_int})
by iteration. In practice we start the iteration by setting $D_{D}=0$
and computing $\omega_{k}$ from (\ref{eq:gkren}) and using this
$\omega_{k}$ and $D_{D}=0$, we compute the perturbed $D_{D}$ using
(\ref{eq:dupre_int}), which can be called $D_{D}^{\left(1\right)}$
since it is the first iteration. Then using this $D_{D}^{\left(1\right)}$
in (\ref{eq:gkren}), we can compute $\omega_{k}^{\left(1\right)}$
and substituting this $\omega_{k}^{\left(1\right)}$ and $D_{D}^{\left(1\right)}$
on the right hand side of (\ref{eq:dupre_int}), we obtain the $D_{D}^{\left(2\right)}$
and so on. We stop the iteration when convergence, defined by $\left|D_{D}^{\left(n\right)}-D_{D}^{\left(n-1\right)}\right|/\left|D_{D}^{\left(n-1\right)}\right|<\epsilon$
where the tolerance $\epsilon$ is taken to be around $\%1$. Fortunately
it takes in general about $3$ iterations for this algorithm to converge.

One issue, which is a common problem in quasi-linear theory is that,
a priori, one can not compute the spectrum $\left|\delta\Phi_{k}\right|^{2}$
that goes into (\ref{eq:dupre_int}) within the theory itself. Renormalization
can be formulated in such a way that we could compute the spectrum
using a local balance condition such as $D_{D}k_{\perp}^{2}\sim\gamma_{k}$.
However while maybe the maximum of the spectrum could be determined
from a condition of this form, the wave-number spectrum in plasma
turbulence is well known to not follow the form implied by $D_{D}k_{\perp}^{2}\sim\gamma_{k}$
at every scale. In particular this is nonsensical when $\gamma_{k}$
becomes negative in some part of the $k$ space. Therefore here we
chose to impose a reasonable form for the $k$-spectrum.
\[
\left|\delta\Phi_{k}\right|^{2}=\begin{cases}
e^{-\left(k-k_{0}\right)^{2}/2\sigma^{2}} & k<k_{p}\\
\left(e^{-\left(k_{p}-k_{0}\right)^{2}/2\sigma^{2}}k_{p}^{3}\right)k^{-3} & k>k_{p}
\end{cases}
\]
where $k_{0}$ is the maximum of the wave-number spectrum around which
it is taken to be a gaussian with a width $\sigma$ which then joins
a power law spectrum at the transition wave-number $k_{p}$. Obviously
the $D_{D}$ that will be computed will depend on these parameters
as well as the plasma parameters $R/L_{n}$, $\eta_{i}$ etc. Note
that the exact form of this power law, or its refinment at higher
$k$ for example by using a spectrum of the form $k^{-3}/\left(1+k^{2}\right)^{2}$
\cite{gurcan:09} makes no practical difference for the computation
of $D_{D}$. Finally we also take $k\rightarrow k_{y}$ above in order
to keep the computation tracktable.

\begin{figure}
\includegraphics[width=1\columnwidth]{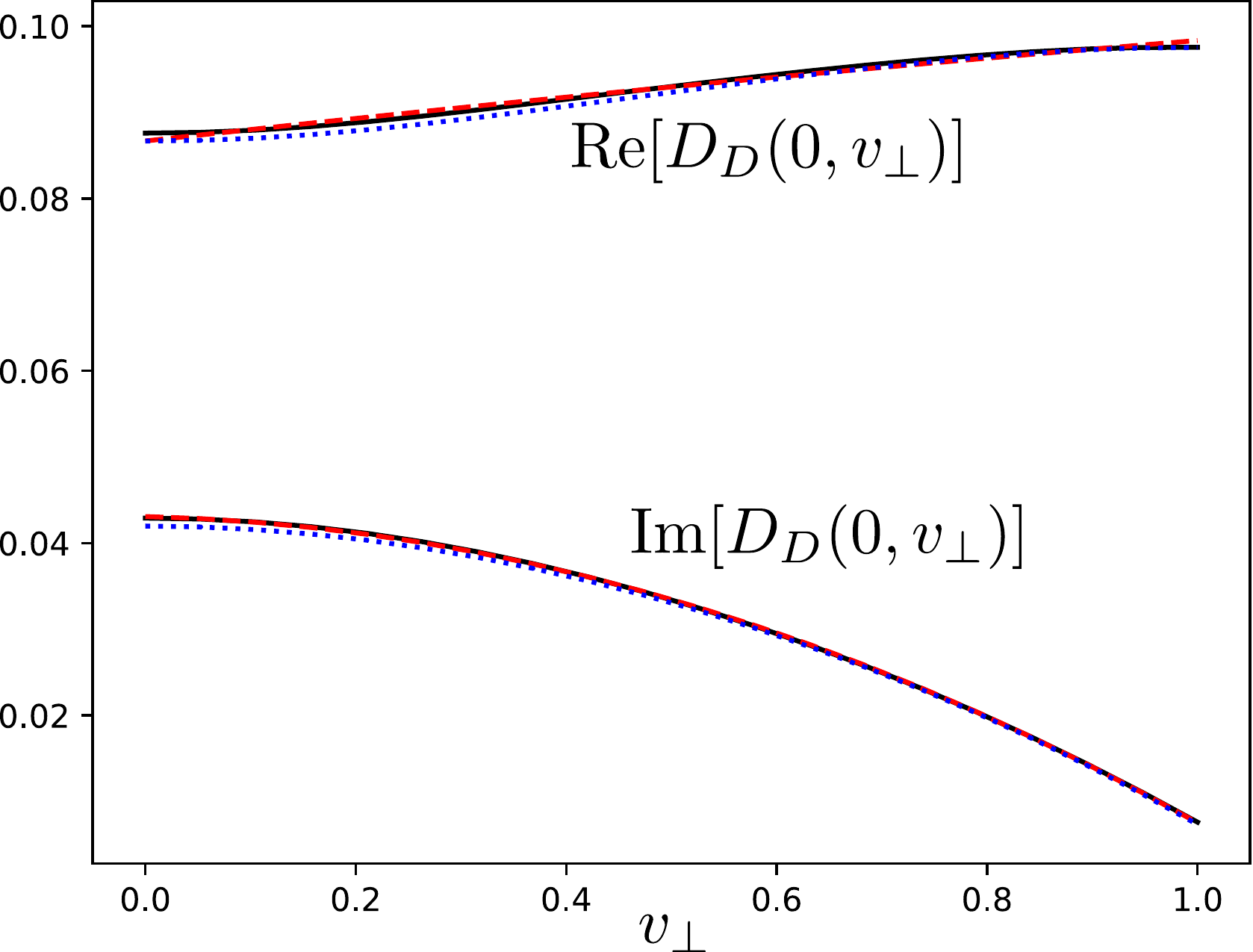}

\caption{\label{fig:ddvperp}The real and imaginary parts of Dupree diffusion
coefficient as a function of $v_{\perp}$. The solid (black) lines
show the $D_{D}(0,v_{\perp})$ computed using iteration at each $v_{\perp}$
by assuming $D_{D}$ is constant in the dispersion relation, dashed
(red) lines show the polynomial fit. The result that is obtained,
when this polynomial fit is used to compute the $D_{D}$ again via
(\ref{eq:drel2}) and (\ref{eq:dupre_int}), is shown as the (blue)
dotted line.}
\end{figure}

\begin{figure}
\includegraphics[width=1\columnwidth]{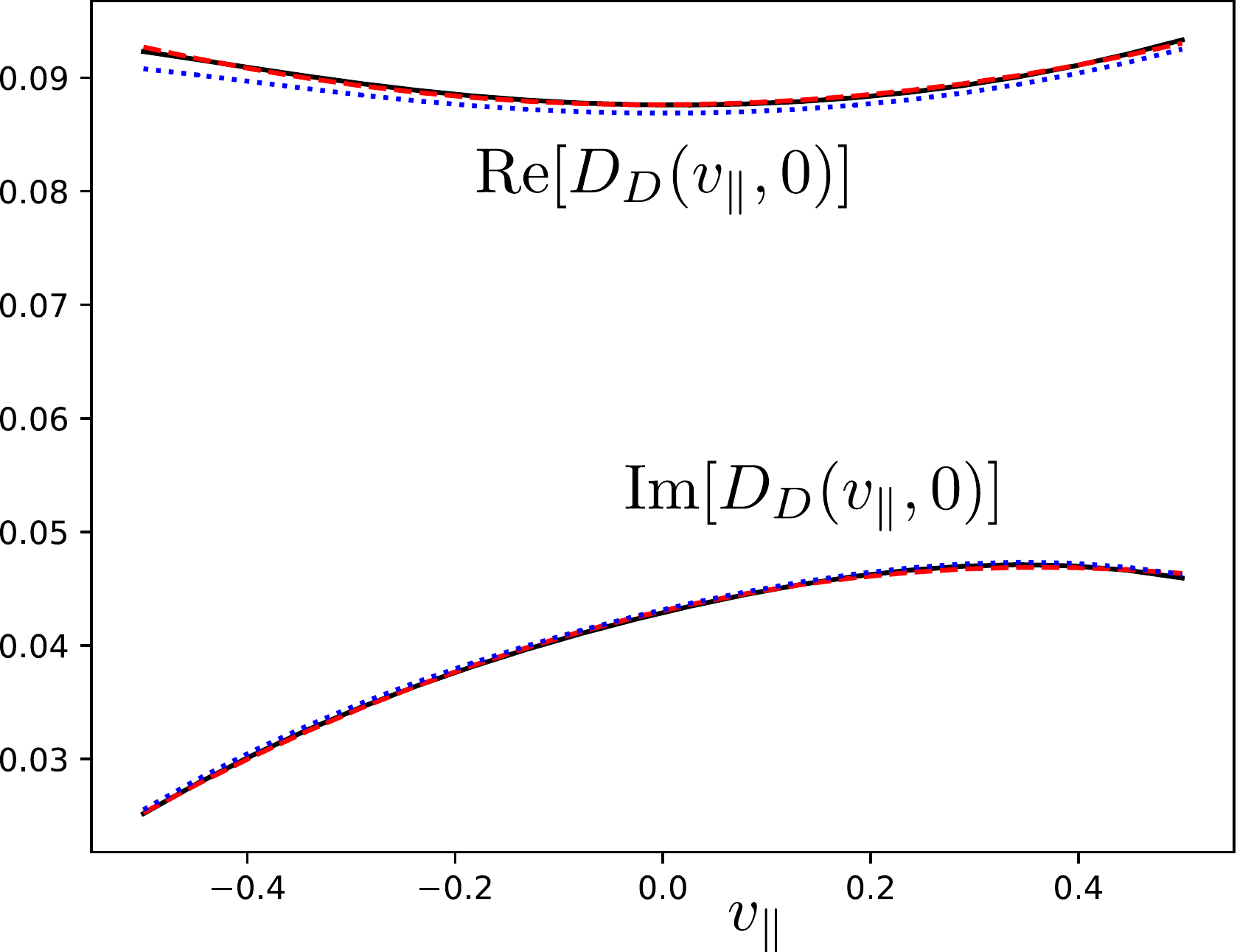}

\caption{\label{fig:ddvpar}The real and imaginary parts of Dupree diffusion
coefficient as a function of $v_{\parallel}$. The solid (black) lines
show the $D_{D}(v_{\parallel},0)$ computed using iteration at each
$v_{\parallel}$ by assuming $D_{D}$ is constant in the dispersion
relation, dashed (red) lines show the polynomial fit. The result that
is obtained, when this polynomial fit is used to compute the $D_{D}$
again via (\ref{eq:drel2}) and (\ref{eq:dupre_int}), is shown as
the (blue) dotted line.}
\end{figure}

With all these assumptions, we can finally compute $D_{D}\left(v\right)$,
the results are shown in figures \ref{fig:ddvperp} and \ref{fig:ddvpar}.
The coefficients of the polynomial (\ref{eq:poly}) are computed in
the two fits, with the parameters $\eta_{i}=2.5$, $L_{n}/R=0.2$,
$k_{\parallel}=0.01$ and $k_{x}=0.1$ as:
\begin{align}
d_{0} & =0.08667974+0.04308331i\nonumber \\
d_{3} & =0.01342785+0.00293719i\nonumber \\
d_{4} & =-0.00176120+0.03260807i\label{eq:coef1}
\end{align}
and 
\begin{align}
d_{0} & =0.08760955+0.04306338i\nonumber \\
d_{1} & =0.00034246+0.02108074i\nonumber \\
d_{2} & =0.02101908+0.02903672i\;\text{.}\label{eq:coef2}
\end{align}
Since the two polynomials are consistent (i.e. $d_{0}$ is approximately
the same in both cases), we can use these coefficients together in
a single two dimensional polynomial form.

\subsection{Solving the renormalized dispersion relation:}

Until this point, we talked about solving the renormalized dispersion
relation (\ref{eq:drel2}). While this is a simple matter of replacing
$\omega\rightarrow\omega+ik_{\perp}^{2}D_{D}$ when $D_{D}$ is taken
to be independent of $v$, when $D_{D}$ is taken to be of the form
(\ref{eq:poly}), the issue is more complicated. In this case, the
dispersion relation have to be rewritten using the $K_{nm}$'s as
follows:
\begin{align}
 & \varepsilon\left(\omega,\mathbf{k}\right)\equiv1+\frac{1}{\tau}+\nonumber \\
 & \frac{1}{\omega_{Di}-id_{2}k_{\perp}^{2}}\bigg[\left(\omega-\omega_{*i}\left(1-\frac{3}{2}\eta_{i}\right)+ik_{\perp}^{2}d_{0}\right)K_{10}\nonumber \\
 & +ik_{\perp}^{2}d_{3}K_{20}+\left(ik_{\perp}^{2}d_{4}-\omega_{*i}\eta_{i}\right)K_{30}\nonumber \\
 & +\left(ik_{\perp}^{2}d_{2}-\omega_{*i}\eta_{i}\right)K_{12}+ik_{\perp}^{2}d_{1}K_{11}\bigg]=0\label{eq:drel3}
\end{align}
where $K_{nm}=K_{nm}\left(\zeta_{a},\zeta_{b},\zeta_{b},\zeta_{c},b\right)$
is used as a shortcut notation with
\[
\zeta_{a}=-\frac{\omega+id_{0}k_{\perp}^{2}}{\omega_{Di}-id_{2}k_{\perp}^{2}}\;\text{,}\quad\zeta_{b}=\frac{\sqrt{2}k_{\parallel}v_{ti}-id_{1}k_{\perp}^{2}}{\omega_{Di}-id_{2}k_{\perp}^{2}}
\]
\[
\zeta_{c}=-\frac{id_{3}k_{\perp}^{2}}{\omega_{Di}-id_{2}k_{\perp}^{2}}\;\text{,}\quad\zeta_{d}=\frac{\frac{1}{2}\omega_{Di}-id_{4}k_{\perp}^{2}}{\omega_{Di}-id_{2}k_{\perp}^{2}}\;\text{.}
\]
When the dispersion relation is written using the $K_{nm}$'s which
are already analytic everywhere on the complex plane thanks to the
analytical continuation discussed above, $\varepsilon\left(\omega,\mathbf{k}\right)$
also becomes analytic everywhere also. Note that the form used in
(\ref{eq:drel2}) is actually not analytical when the imaginary part
of $\rho_{+}$ as defined in (\ref{eq:phopm}) is greater than zero.
Using a least square solver, we can then find the zeros of the renormalized
dispersion relation and trace these results. The result with combined
set of coefficients from (\ref{eq:coef1}) and (\ref{eq:coef2}) is
shown in figure 

\begin{figure}
\includegraphics[width=1\columnwidth]{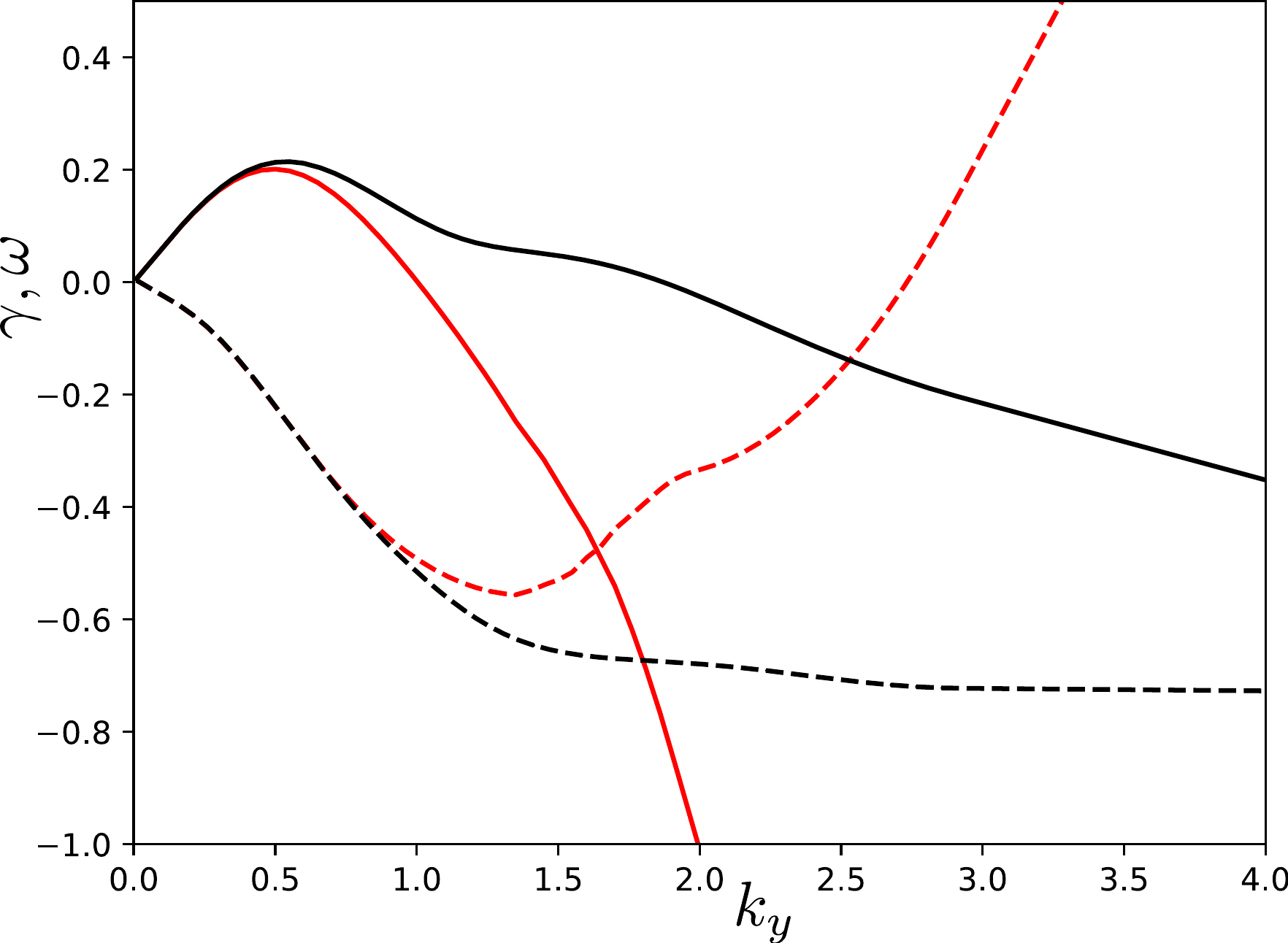}

\caption{The growth rate (solid lines) and the frequency (dashed lines), with
(red) and without (black) renormalization. Note that while the frequency
changes sign and becomes positive, since the growth rate becomes strongly
negative, it is unlikely that this positive frequency can actually
be realized.}

\end{figure}

\section{Results and Conclusions}

Using a generalization of curvature modified plasma dispersion functions,
we were able to implement a linear solver that can solve the dispersion
relation with a Dupree diffusion coefficient in the form of a second
order polynomial. The resulting solver was used in an algorithm based
on iteration in order to solve the Dupree integral relation and therefore
obtain the renormalized Dupree diffusion coefficient together with
the renormalized growth rate and frequency for the gyrokinetic, local,
electrostatic ITG, with adiabatic electrons.

Since the growth rate is obtained as the solution of the renormalized
dispersion relation, it becomes rapidly negative for larg $k$ , in particular, due to the effect of
Dupree diffusion. Thus, one has to define the generalized curvature modified
plasma dispersion functions, the $K_{nm}$'s together with their analytical
continuation. This guarantees that when the dispersion relation is
written as in (\ref{eq:drel3}) using the $K_{nm}$'s it is also analytic
everywhere in the complex plane.

The Dupree diffusion coefficient is rather important in quasi-linear
transport formulation\cite{bourdelle:07}, which has been developed into a full transport framework over the years \cite{citrin:17}, since in this formulation the effective correlation time is argued to be renormalized as compared to the linear one. It is also true that the heat flux in ITG should be proportional to the
Dupree diffusion coefficient\cite{balescu:book:anom}.

The basic computation that is shown here is based on a number of
assumptions such as a particular form that one has to assume for the
$k$-spectrum, or the assumption that during the iteration procedure
for every $v$ value one can at first assume that $D_{D}$ is a constant,
etc. Nonetheless it proposes a numerically tractable, practical renormalization
algorithm based on iteration including curvature effects. The algorithm can be generalized to
include $k$ dependence of the eddy damping as well using a closure
such as direct interaction approximation, or the eddy damped quasi-normal
Markovian approximation, or rather its realizable variant\cite{lesieur:book,krommes:02}.

\end{document}